\documentclass{article}

\usepackage{microtype}
\usepackage{graphicx}
\graphicspath{{images/}}
\usepackage{booktabs} 
\usepackage{cite}
\usepackage{mathtools}
\usepackage{amsmath,amssymb,amsfonts}
\usepackage{algorithmic}
\usepackage{textcomp}
\usepackage{pstricks}
\usepackage{mathdots}
\usepackage{subcaption}

\usepackage{hyperref}

\usepackage[accepted]{icml2018}

\icmltitlerunning{Network Lens}

\begin{document}

\twocolumn[
\icmltitle{Network Lens: Node Classification in\\Topologically Heterogeneous Networks}




\begin{icmlauthorlist}
\icmlauthor{Kshiteesh Hegde}{rpi}
\icmlauthor{Malik Magdon-Ismail}{rpi}
\end{icmlauthorlist}

\icmlaffiliation{rpi}{Department of Computer Science, Rensselaer Polytechnic Institute, Troy, NY, USA}

\icmlcorrespondingauthor{Kshiteesh Hegde}{hegdek2@cs.rpi.edu}
\icmlcorrespondingauthor{Malik Magdon-Ismail}{magdon@cs.rpi.edu}

\icmlkeywords{graph classification, network lens, deep learning, network signature}

\vskip 0.3in
]



\printAffiliationsAndNotice{} 

\begin{abstract}
We study the problem of identifying different behaviors occurring in different parts of a large heterogenous network. We \emph{zoom} in to the network using \emph{lenses} of different sizes to capture the local structure of the network. These network \emph{signatures} are then weighted to provide a set of predicted labels for every node. We achieve a peak accuracy of $\sim$42\% (random=11\%) on two networks with $\sim$100,000 and $\sim$1,000,000 nodes each. Further, we perform better than random even when the given node is connected to up to 5 different types of networks. Finally, we perform this analysis on homogeneous networks and show that highly structured networks have high homogeneity.
\end{abstract}

\section{Introduction}
\label{sec:intro}
\begin{figure}
\centering
\includegraphics[scale=0.6]{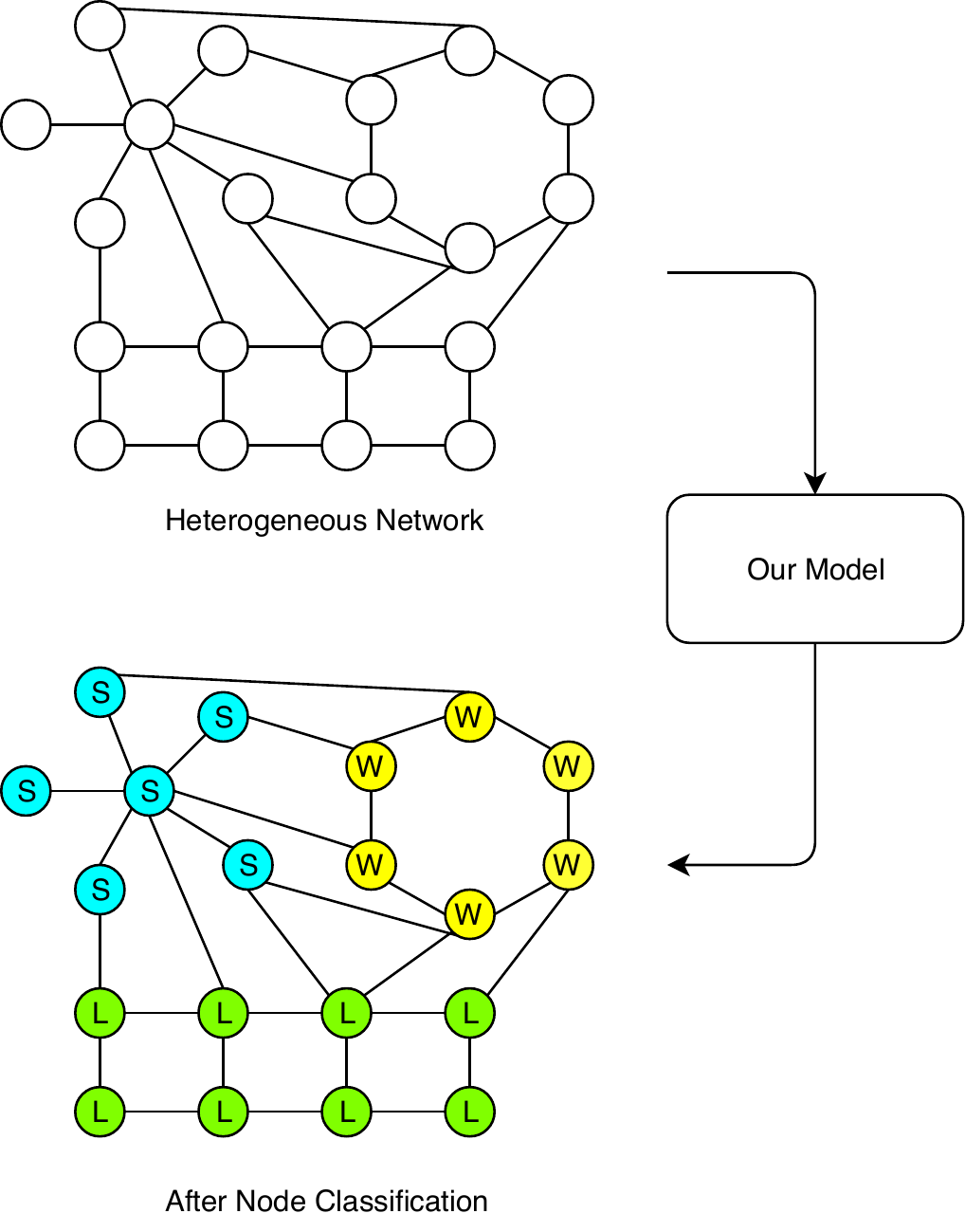}
\caption{Example of an application of our model: Clustering}
\label{fig:example}
\end{figure}

Large networks, which are a direct result of the ever expanding big data world, are commonplace in almost every domain. Social networks such as Facebook and Twitter, customer purchase data on e-commerce platforms like amazon.com, author citation records like DBLP, road networks in any country/state are all popular examples of large networks. In a heterogeneous setting, one could be dealing with the situation where parts of a network are behaving differently. For example, in a social network, some parts may behave collaboratively, some may be terrorist-like, etc. Conventional graph classification approaches will fail to notice such differences in a network. Identifying the behavior of different parts of a network is a key problem. One of the applications of solving this node classification problem is clustering. We illustrate this in Figure \ref{fig:example}.

We start with a network that potentially is exhibiting different behaviors in different parts. The model takes in subgraphs of different sizes using random walks depending on the \emph{lens size} from various parts of the network. We call it the \emph{lens} since one can do a random walk starting from any node in the network to capture local structure in that region much similar to how one can hover a metaphorical lens on any area of the network to zoom in and see that area in more detail.

Then, a set of labels with associated probabilities (confidence of the model) is outputted. We can visualize this in the colored graph in Figure \ref{fig:example} which illustrates the output on a toy example. The model has identified three different behaviors in the network: a star graph, a wheel graph and a ladder graph denoted by cyan (S), yellow (W) and green (L) respectively. Thus, our node classification model has clustered the given graph into three differently behaving parts. Clustering is just one of the potential applications of our node classification lens.

\begin{figure}
\psscalebox{0.9 0.9} 
{
\begin{pspicture}(0, 3)(9, 14)
\rput(4,11.5){\includegraphics[scale=0.25]{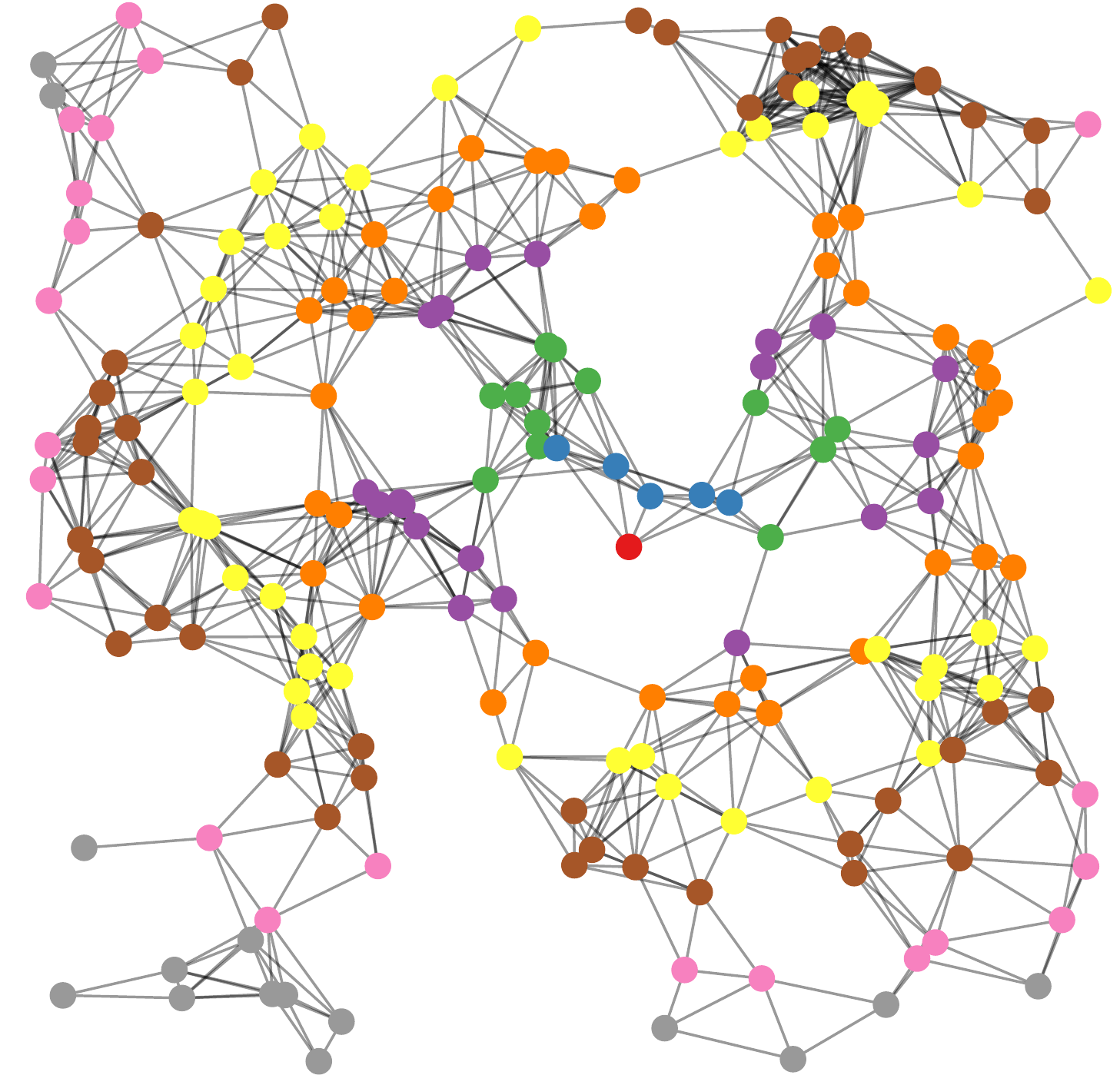}}
\rput(7.5, 11.5){\shortstack{Heterogenous\\Network}}

\psline[linecolor=black, linewidth=0.01, linestyle=solid](2.45,12.5)(1.4,12.5)
\rput(1, 12.55){Lens}
\pscircle[linecolor=black, linewidth=0.08, linestyle=solid, dimen=outer, strokeopacity=0.5](2.8, 12.5){0.35}
\pscircle[linecolor=black, linewidth=0.08, linestyle=solid, dimen=outer, strokeopacity=0.5](5.2, 10.5){0.35}
\psline[linecolor=black, linewidth=0.04, linestyle=solid, arrowsize=0.05291667cm 2.0,arrowlength=1.4,arrowinset=0.0]{->}(2.8,12.15)(2.8,8.8)
\psline[linecolor=black, linewidth=0.04, linestyle=solid, arrowsize=0.05291667cm 2.0,arrowlength=1.4,arrowinset=0.0]{->}(5.2,10.15)(5.2,8.8)

\rput(2.8,7.8){\fboxsep0pt\fbox{\includegraphics[scale=0.25]{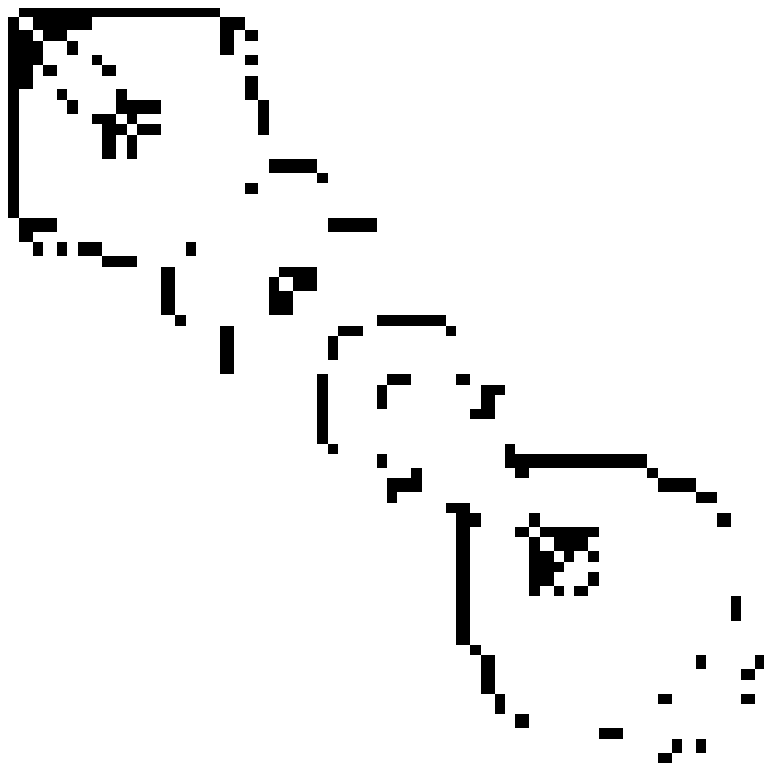}}}
\rput(5.2,7.8){\fboxsep0pt\fbox{\includegraphics[scale=0.25]{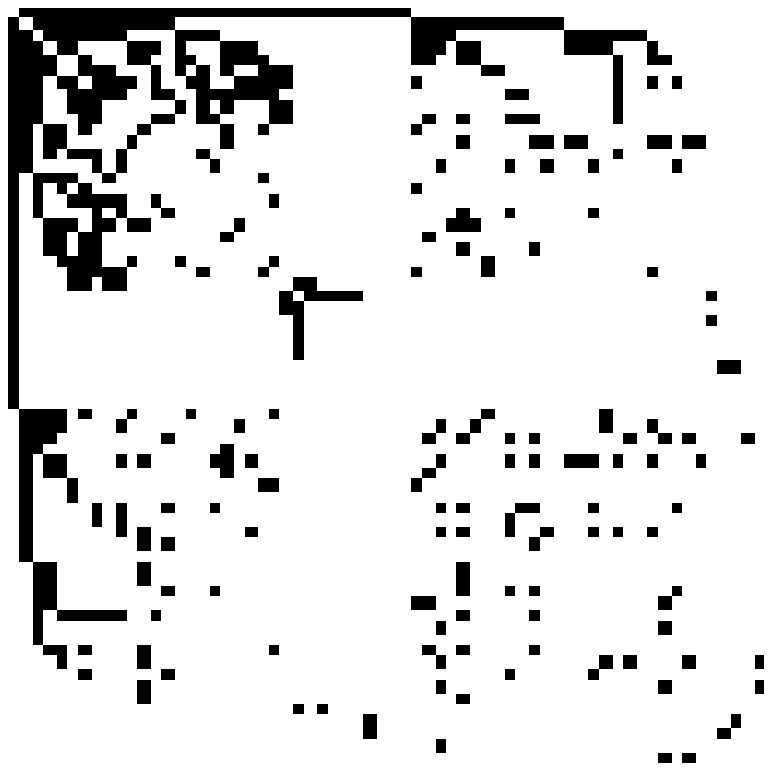}}}
\rput(7.5, 7.8){\shortstack{Network\\Signatures}}

\psline[linecolor=black, linewidth=0.04, arrowsize=0.05291667cm 2.0,arrowlength=1.4,arrowinset=0.0]{->}(2.8, 6.8)(2.8, 5)
\psline[linecolor=black, linewidth=0.04, arrowsize=0.05291667cm 2.0,arrowlength=1.4,arrowinset=0.0]{->}(5.2, 6.8)(5.2, 5)

\psframe[linecolor=black, linewidth=0.04, dimen=outer](1.8, 5)(6.2, 4)
\rput(4, 4.5){Classify}
\end{pspicture}
}
\caption{Workflow of Network Lens. Image embeddings of local adjacency matrices are the input to a classifier, which produces node labels.}
\label{fig:outline}
\end{figure}

We study the general problem: Given a large, potentially heterogenous, network, can one identify the different behaviors occurring in the network? The outline of our approach to the problem is shown in Figure \ref{fig:outline}.

\begin{figure}
\begin{subfigure}{0.25\textwidth}
	\centering
	\fboxsep0pt\fbox{\includegraphics[scale=0.5]{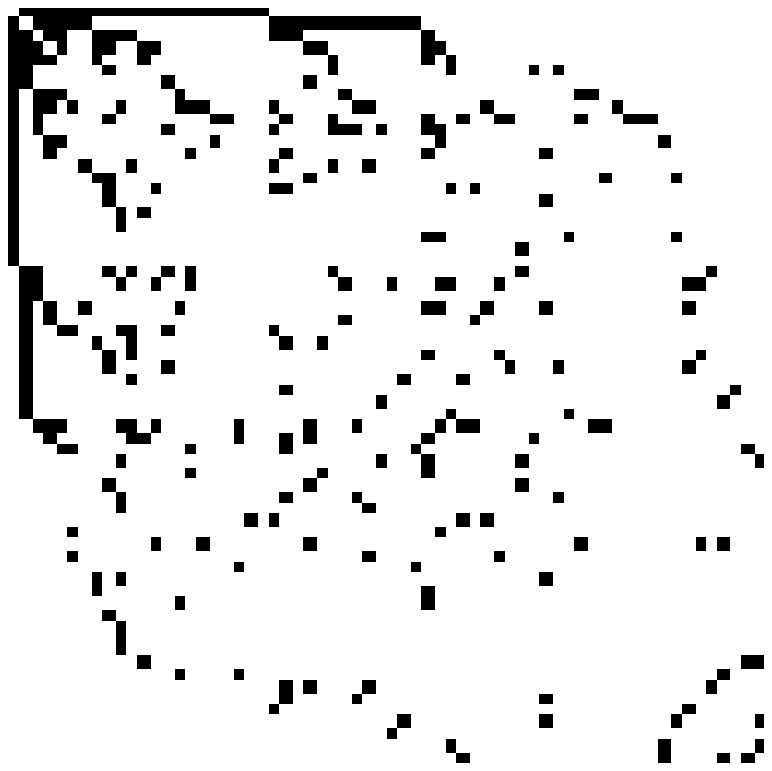}}
	\caption{Wikipedia}
	\label{fig:wikipedia}
\end{subfigure}%
\begin{subfigure}{0.25\textwidth}
	\centering
	\fboxsep0pt\fbox{\includegraphics[scale=0.5]{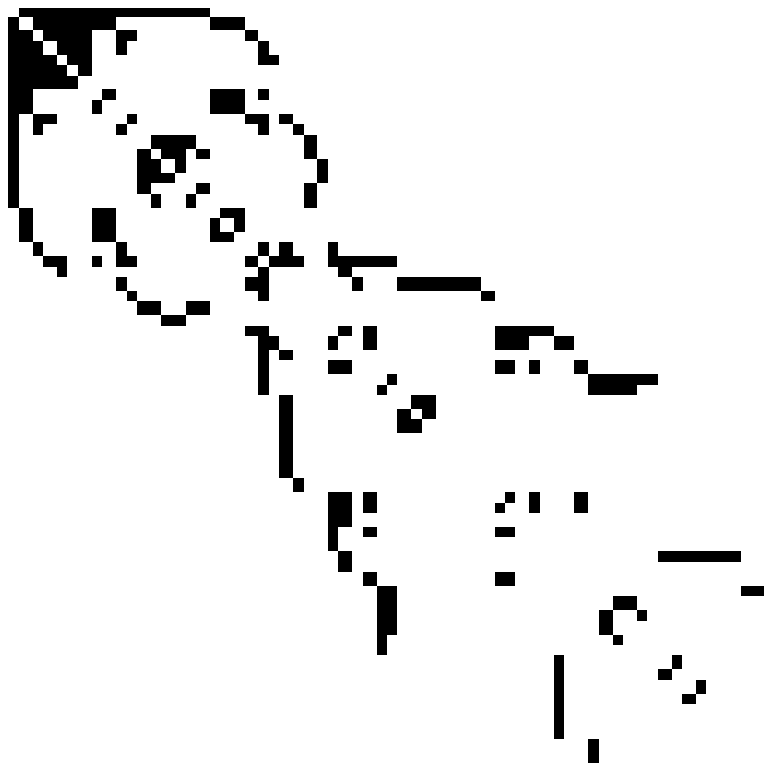}}
	\caption{Terrorist Network}
	\label{fig:aq}
\end{subfigure}
\begin{subfigure}{0.25\textwidth}
	\centering
	\fboxsep0pt\fbox{\includegraphics[scale=0.5]{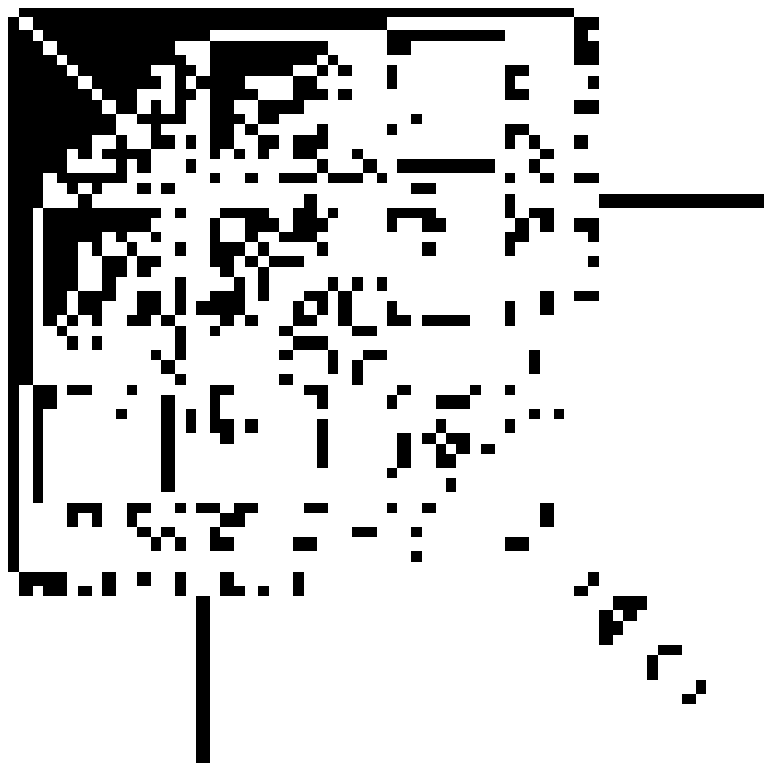}}
	\caption{Facebook}
	\label{fig:facebook}
\end{subfigure}%
\begin{subfigure}{0.25\textwidth}
	\centering
	\fboxsep0pt\fbox{\includegraphics[scale=0.5]{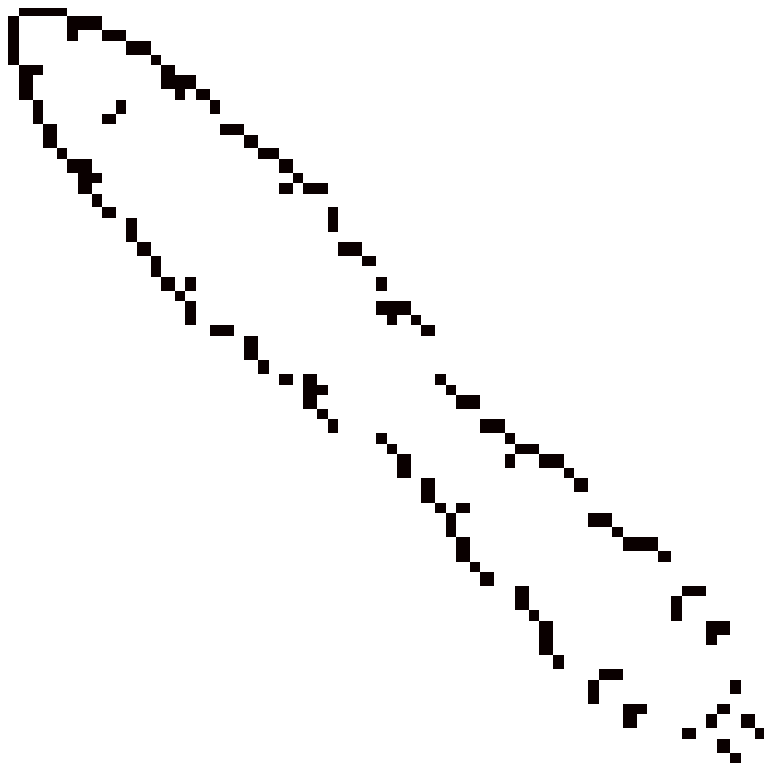}}
	\caption{Road Network}
	\label{fig:roadnet}
\end{subfigure}
\caption{Signatures of select networks to demonstrate the structured image embedding feature.}
\label{fig:sigs}
\end{figure}

The network signatures shown in Figure \ref{fig:sigs} as introduced in \cite{wu2016asonam} are the structured adjacency matrices of subgraphs picked up by lenses of different sizes. In the image, a black pixel at position ($i$, $j$) denotes an edge between nodes $i$ and $j$. It is structured according to the ordering scheme presented in \cite{wu2016asonam} and is presented in more detail in Section \ref{sec:imem}. They are a powerful representation \cite{hegdesig2018} of networks since they are agnostic to the type of the network. They can be applied to a wide variety of networks including, but not limited to, social, information, transportation and even terrorist networks. One of the applications of this representation is subnetwork classification. They are good for machine learning algorithms and have an intuitive visual representation.

A signature of a network comes primarily from its function. For example, the function of a road network is transportation. The functions of a transportation network include having the ability to connect local places in a city as well as distant cities via highways. It also needs to manage traffic during rush hours. It need not have connections from every place to every other place but the it has a large cut. The networks that evolve to support a transportation function are grid-like. That is the signature of a road network (Figure \ref{fig:roadnet}). The connections in a network evolve in a particular way to support a particular function and develop a signature. Similarly, different networks with different functions evolve in different ways and develop different signatures. 

In this work, we use the signatures to build a network lens. We use models previously trained on these signatures (see Section \ref{sec:datamethod}) from the homogenous setting to obtain a set of predicted labels from each of the lenses for each node. We then use linear programming to arrive at the optimal weights for labels of different lenses and construct a final set of predicted labels for each node.

To solve the general problem of node classification via network lens, we need new ways to:
\begin{enumerate}
	\item Classify a small subnetwork of a network into one of several types
	\item Test the accuracy of such algorithms
	\item Evaluate performance on real networks
\end{enumerate}

\subsection*{Our Contributions}
\begin{enumerate}
\item We transform the problem of graph classification in to one of image classification. We show that even at a tiny local scale of up to 8 nodes we can classify nodes in a heterogenous network with $\sim1,000,000$ nodes and achieve accuracies up to 42\% which is well above random performance of 11\%.

\item We note that when a node is more diverse (having multiple connections to different types of networks), it is harder to predict that node's type correctly. However, our lossless image feature is powerful enough that even when a node is connected to up to 5 different types of networks, we perform better than random. In the real world, where a node with two types of connections is the most common scenario, our technique is significantly better than random with $\sim$32\% accuracy.

\item Finally, we test our model on real networks to study their degree of heterogeneity. We find that some networks are highly homogenous whereas others have a high degree of heterogeneity.
\end{enumerate}

\section{Related Work}
\label{sec:relwork}
We study the problem of identifying different behaviors in a network by using image classification to categorize local structure.

The idea of using the image embedding of the adjacency matrix as a feature was first introduced in \cite{wu2016asonam}. Based on this idea, authors in \cite{hegdesig2018} showed with great success that parent networks of tiny subgraphs (as small as 8 nodes) can be identified. They also used Caffe \cite{jia2014caffe} to show that the structured image embedding features can be used for classification in a transfer learning setting. In this work, we use the idea to create a \emph{lens} that can be used on heterogeneous networks to \emph{see} the different \emph{behaviors} exhibited in different parts of a network.

The most popular approaches to graph classification are feature selection and kernel methods. Authors in \cite{kong2010} perform semi-supervised feature selection by searching for optimal subgraph features. They define a metric that governs how features are selected. There is also the idea of using pattern recognition along with feature selection where the idea is that graphs from the same class should have similar attributes \cite{geng2012}. Spatial distribution of subgraphs is used as features in \cite{fei2008}. In a similar vein, \cite{jin2009} introduces a pattern exploration scheme that looks for co-occurring features in subgraphs to perform binary classification. It is unclear how multi-class classification can be achieved (if at all) using this approach. In \cite{kong2010multi}, the authors talk about extracting important features in a multi-label setting. They assume that the given data is already labeled (multiple times) and the task is to choose the correct label from the set. All the above methods require construction of features that are dependent on the given data. This can be non-trivial in cases where one has to deal with a diverse set of data as is the case in this study. Developing a one size fits all kind of a set of features is near impossible. In case of pattern recognition, if a new pattern or set of patterns emerge only in the test set, then the chances of catching them drastically decreases.

Many graph kernels based on walks, subtrees, cycles, shortest paths etc. have been proposed \cite{kernelsKashima2002,marginalizedKashima2003,boostKudo2004,borgwardt2005shortest,gartner2003graph,riesen2009graph}. The kernel function computes the similarity between two graphs and then a classifier such as SVM is used for classification. As evidenced by the abundance of different types of kernel functions, it is difficult to come up with a kernel that ticks all the boxes for a given classification problem. The size and domain of the network, complexity of the kernel function all affect the decision of choosing the \emph{right} kernel. So, kernel methods are also affected by the same problems as feature selection methods.

All of the above mentioned literature assume a friendly setting where one network contains only one type of network. They are of little use when different types of subgraphs are connected to each other in the same network. This amounts to different parts \emph{behaving} differently. This setting is much more difficult than the friendly setting as we demonstrate later.

The lossless structured image embedding feature used in this work, solves the above problems. It focuses on the structure that networks exhibit at a local level independent of the domain of origin of the network. As we show later, this approach works even when different classes of subgraphs are connected to each other in the same network. Since, we could not find similar methods introduced by previous researchers, we believe this is a significant result in the field of heterogenous node classification.

\section{Data and Methodology}
\label{sec:datamethod}

\subsection{Data}
\label{sec:data}
We used 9 real world networks to construct our heterogenous network. The networks are from a diverse set of domains like e-commerce, social, web, roads etc. Table \ref{tab:data} provides the number of nodes and edges in each of the individual networks.

\begin{table*}[h]
\centering
\resizebox{0.9\textwidth}{!}{%
\begin{tabular}{c|c|c}
Dataset & \# Nodes & \# Edges\tabularnewline
\hline 
Citation \cite{citationLeskovec2005,citationGehrke2003} & 34,546 & 421,578\tabularnewline
Facebook \cite{facebookMcAuley2012} & 4039 & 88,234\tabularnewline
Road Network \cite{roadLeskovec2009} & 1,088,092 & 1,541,898\tabularnewline
Web \cite{roadLeskovec2009} & 875,713 & 5,105,039\tabularnewline
Wikipedia \cite{wikiWest2012,wiki2West2009} & 4,604 & 119,882\tabularnewline
Amazon \cite{amazonLeskovec2007} & 334,863 & 925,872\tabularnewline
DBLP \cite{dblpYang2012} & 317,080 & 1,049,866\tabularnewline
Terrorist Net. \cite{jjatt} & 271 & 756\tabularnewline
Gowalla \cite{cho2011friendship} & 196,591 & 950,327\tabularnewline
\end{tabular}}
\caption{The homogenous networks used in this study}
\label{tab:data}
\end{table*}

\subsection{Construction of Heterogenous Networks}
\label{sec:consthn}
Our first task is to construct a heterogenous testbed using real networks. Each of the real world networks in Table \ref{tab:data} behaves differently and has a different individual signature \cite{hegdesig2018}. We take several snapshots of each of these networks and splice them together to obtain one big heterogeneous network. This ensures that different regions of the spliced network possess different local signatures.

\begin{figure}

\psscalebox{0.8 0.8} 
{
\begin{pspicture}(0,-7.3598633)(7.899275,3.3598633)
\psframe[linecolor=black, linewidth=0.04, dimen=outer](4.749637,3.3598633)(3.1496372,2.1598632)
\psframe[linecolor=black, linewidth=0.04, dimen=outer](7.5496373,3.3598633)(5.9496374,2.1598632)
\psline[linecolor=black, linewidth=0.04, arrowsize=0.05291667cm 2.0,arrowlength=1.4,arrowinset=0.0]{<-}(1.5496373,-0.6401367)(3.1496372,2.1598632)
\psline[linecolor=black, linewidth=0.04, arrowsize=0.05291667cm 2.0,arrowlength=1.4,arrowinset=0.0]{<-}(6.3496375,-0.6401367)(4.749637,2.1598632)
\psline[linecolor=black, linewidth=0.04, arrowsize=0.05291667cm 2.0,arrowlength=1.4,arrowinset=0.0]{<-}(3.1496372,-0.6401367)(3.9496374,2.1598632)
\psline[linecolor=black, linewidth=0.04, arrowsize=0.05291667cm 2.0,arrowlength=1.4,arrowinset=0.0]{<-}(4.749637,-0.6401367)(3.9496374,2.1598632)(3.9496374,2.1598632)
\psframe[linecolor=black, linewidth=0.04, dimen=outer](1.9496373,-0.6401367)(1.1496373,-1.4401367)
\psframe[linecolor=black, linewidth=0.04, dimen=outer](3.5496373,-0.6401367)(2.7496374,-1.4401367)
\psframe[linecolor=black, linewidth=0.04, dimen=outer](5.149637,-0.6401367)(4.3496375,-1.4401367)
\psframe[linecolor=black, linewidth=0.04, dimen=outer](6.749637,-0.6401367)(5.9496374,-1.4401367)
\psline[linecolor=black, linewidth=0.04, linestyle=dotted, dotsep=0.10583334cm, arrowsize=0.05291667cm 2.0,arrowlength=1.4,arrowinset=0.0]{->}(5.9496374,2.1598632)(5.5496373,1.3598633)
\psline[linecolor=black, linewidth=0.04, linestyle=dotted, dotsep=0.10583334cm, arrowsize=0.05291667cm 2.0,arrowlength=1.4,arrowinset=0.0]{->}(6.749637,2.1598632)(6.3496375,1.3598633)
\psline[linecolor=black, linewidth=0.04, linestyle=dotted, dotsep=0.10583334cm, arrowsize=0.05291667cm 2.0,arrowlength=1.4,arrowinset=0.0]{->}(6.749637,2.1598632)(7.149637,1.3598633)
\psline[linecolor=black, linewidth=0.04, linestyle=dotted, dotsep=0.10583334cm, arrowsize=0.05291667cm 2.0,arrowlength=1.4,arrowinset=0.0]{->}(7.5496373,2.1598632)(7.9496374,1.3598633)
\psline[linecolor=black, linewidth=0.04, linestyle=dotted, dotsep=0.10583334cm](2.3496373,2.8098633)(2.7496374,2.8098633)(2.7496374,2.8098633)
\psline[linecolor=black, linewidth=0.04, linestyle=dotted, dotsep=0.10583334cm](5.149637,2.8098633)(5.5496373,2.8098633)
\psframe[linecolor=black, linewidth=0.04, dimen=outer](1.9496373,3.3598633)(0.3496373,2.1598632)
\psline[linecolor=black, linewidth=0.04, linestyle=dotted, dotsep=0.10583334cm, arrowsize=0.05291667cm 2.0,arrowlength=1.4,arrowinset=0.0]{->}(0.3496373,2.1598632)(-0.050362702,1.3598633)
\psline[linecolor=black, linewidth=0.04, linestyle=dotted, dotsep=0.10583334cm, arrowsize=0.05291667cm 2.0,arrowlength=1.4,arrowinset=0.0]{->}(1.1496373,2.1598632)(0.7496373,1.3598633)
\psline[linecolor=black, linewidth=0.04, linestyle=dotted, dotsep=0.10583334cm, arrowsize=0.05291667cm 2.0,arrowlength=1.4,arrowinset=0.0]{->}(1.1496373,2.1598632)(1.5496373,1.3598633)
\psline[linecolor=black, linewidth=0.04, linestyle=dotted, dotsep=0.10583334cm, arrowsize=0.05291667cm 2.0,arrowlength=1.4,arrowinset=0.0]{->}(1.9496373,2.1598632)(2.3496373,1.3598633)
\psline[linecolor=black, linewidth=0.04, linestyle=dotted, dotsep=0.10583334cm](1.1496373,0.9598633)(1.1496373,-0.24013671)
\psline[linecolor=black, linewidth=0.04, linestyle=dotted, dotsep=0.10583334cm](6.749637,0.9598633)(6.749637,-0.24013671)(6.749637,-0.24013671)
\psline[linecolor=black, linewidth=0.04, linestyle=dotted, dotsep=0.10583334cm, arrowsize=0.05291667cm 2.0,arrowlength=1.4,arrowinset=0.0]{->}(1.5496373,-1.4401367)(2.3496373,-3.4401367)
\psline[linecolor=black, linewidth=0.04, linestyle=dotted, dotsep=0.10583334cm, arrowsize=0.05291667cm 2.0,arrowlength=1.4,arrowinset=0.0]{->}(3.1496372,-1.4401367)(3.9496374,-3.4401367)
\psline[linecolor=black, linewidth=0.04, linestyle=dotted, dotsep=0.10583334cm, arrowsize=0.05291667cm 2.0,arrowlength=1.4,arrowinset=0.0]{->}(4.749637,-1.4401367)(3.9496374,-3.4401367)
\psline[linecolor=black, linewidth=0.04, linestyle=dotted, dotsep=0.10583334cm, arrowsize=0.05291667cm 2.0,arrowlength=1.4,arrowinset=0.0]{->}(6.3496375,-1.4401367)(5.5496373,-3.4401367)

\rput(3.9496374, -5.4401367){\includegraphics[scale=0.25]{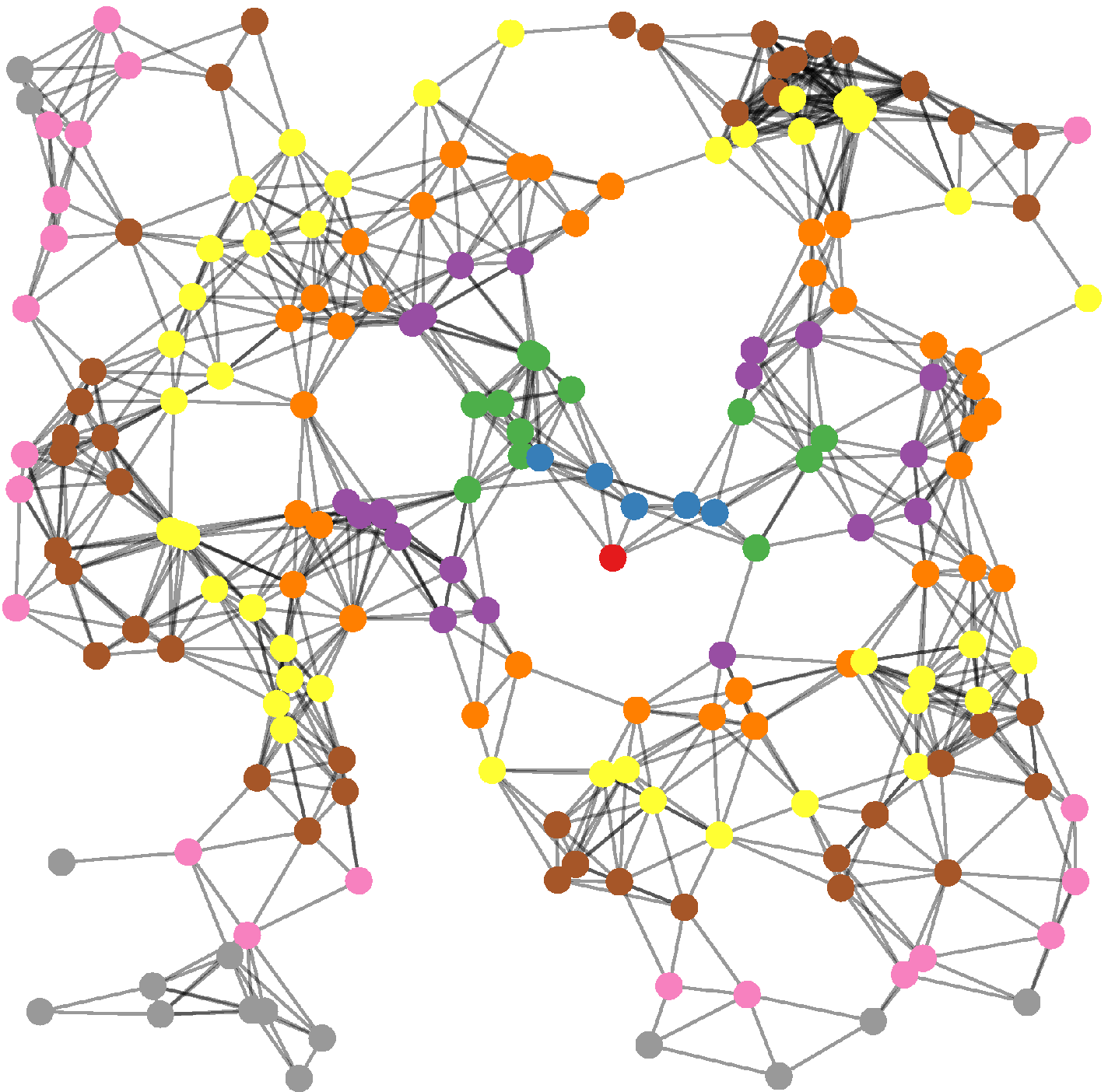}}

\rput(1.1596373,2.8098632){Amazon}
\rput(3.9096373,2.8098632){DBLP}
\rput(6.739637,2.8098632){Wiki}
\rput(1.5496373,-1.0401367){8}
\rput(3.1496372,-1.0401367){16}
\rput(4.749637,-1.0401367){32}
\rput(6.3496375,-1.0401367){64}
\rput(7.6496375, -1.0401367){\shortstack{Subgraph\\size}}
\rput{72}(3.14, 0.8){Extract subgraphs}
\end{pspicture}
}

\caption{Construction of the heterogenous networks by splicing together subgraphs from real networks.}
\label{fig:consthn}
\end{figure}

First, we extract 100 subgraphs with 8, 16, 32 and 64 nodes from each of the networks shown in Table \ref{tab:data} yielding 3600 (100 $\times$ 4 $\times$ 9) disjoint subgraphs in total. Next, we choose a pair of subgraphs at random and choose a node from each of these two subgraphs randomly and introduce an edge between them. This edge-introduction process is repeated 10 $\times$ 3600 times resulting in a connected heterogenous network with 108,000 nodes and 294,841 edges. We constructed a bigger heterogenous network similarly, but with 1000 subgraphs resulting in a network with 1,080,000 nodes and 2,951,234 edges. This process in illustrated in Figure \ref{fig:consthn}.

\subsection{Graph Image Embeddings}
\label{sec:imem}
\begin{figure}

\psscalebox{0.9 0.9} 
{
\begin{pspicture}(-1.7,-4.503479)(5.24,5.353479)
\rput(0.02, 4.35){\includegraphics[scale=0.25]{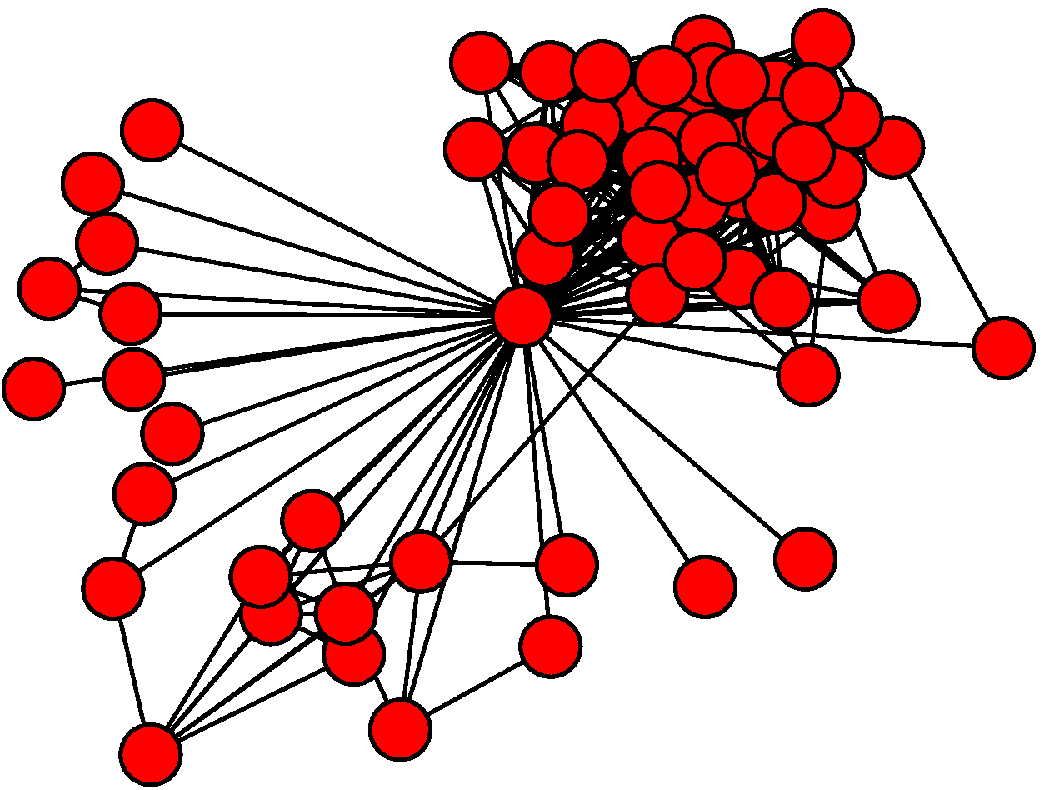}}
\psline[linecolor=black, linewidth=0.04, arrowsize=0.05291667cm 2.0,arrowlength=1.4,arrowinset=0.0]{->}(0.02,3.353479)(0.02,1.753479)
\rput(5.22, 4.35){\includegraphics[scale=0.25]{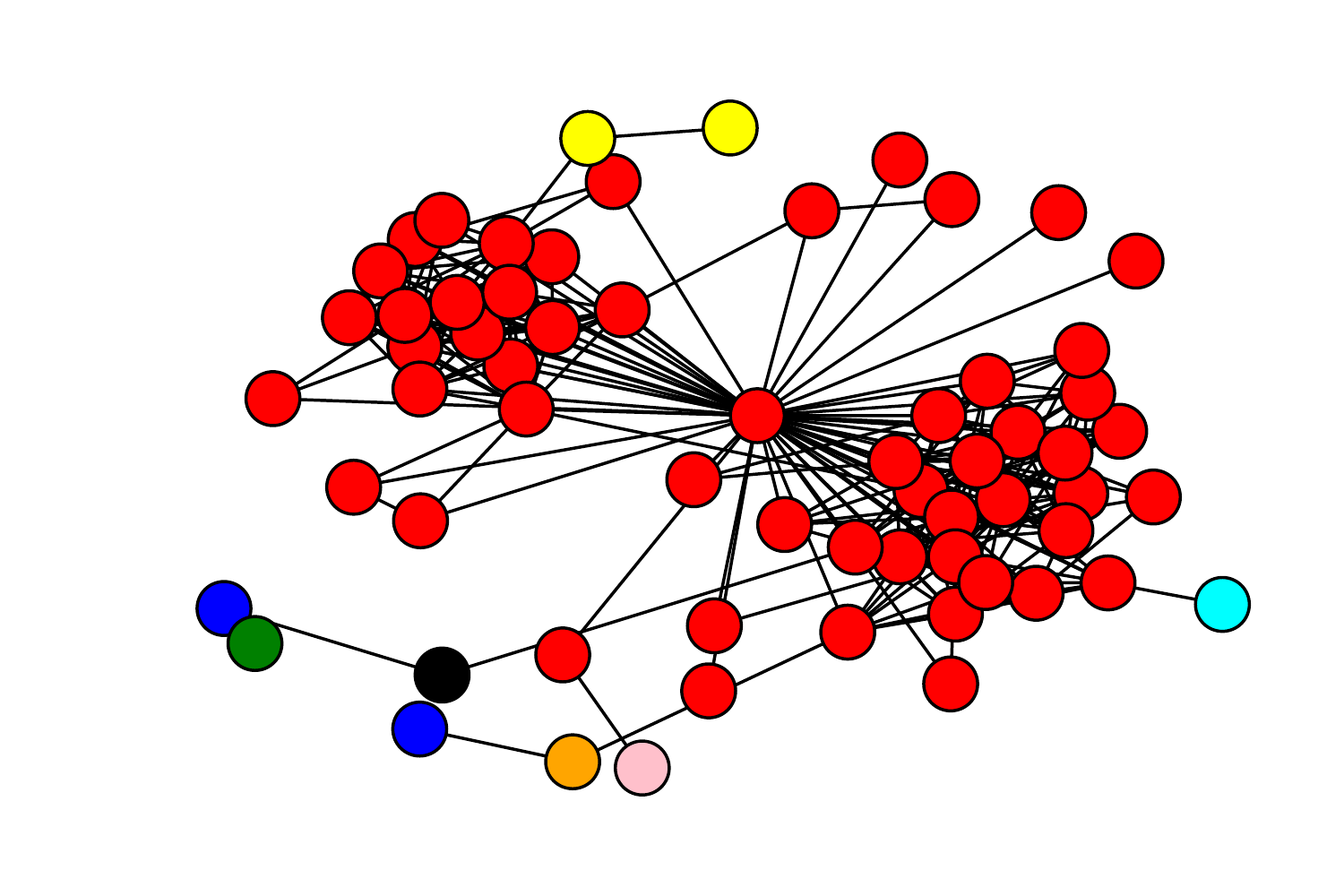}}
\psline[linecolor=black, linewidth=0.04, arrowsize=0.05291667cm 2.0,arrowlength=1.4,arrowinset=0.0]{->}(5.22,3.353479)(5.22,1.753479)
\rput(0.02, 0.75){\fboxsep0pt\fbox{\includegraphics[scale=0.25]{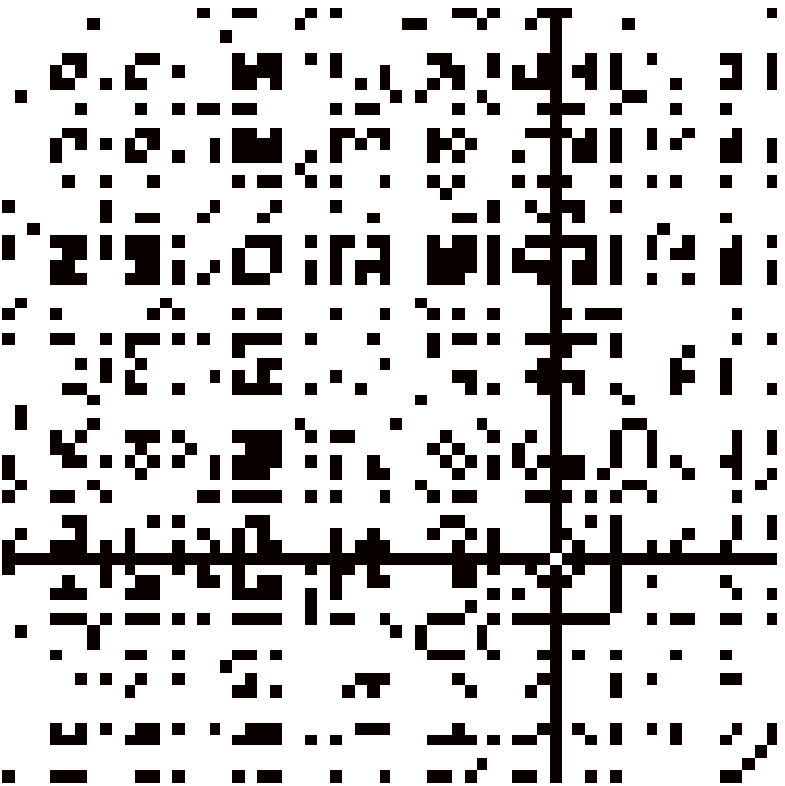}}}
\psline[linecolor=black, linewidth=0.04, arrowsize=0.05291667cm 2.0,arrowlength=1.4,arrowinset=0.0]{->}(0.02,-0.406521)(0.02,-2.006521)
\rput(5.22, 0.75){\fboxsep0pt\fbox{\includegraphics[scale=0.25]{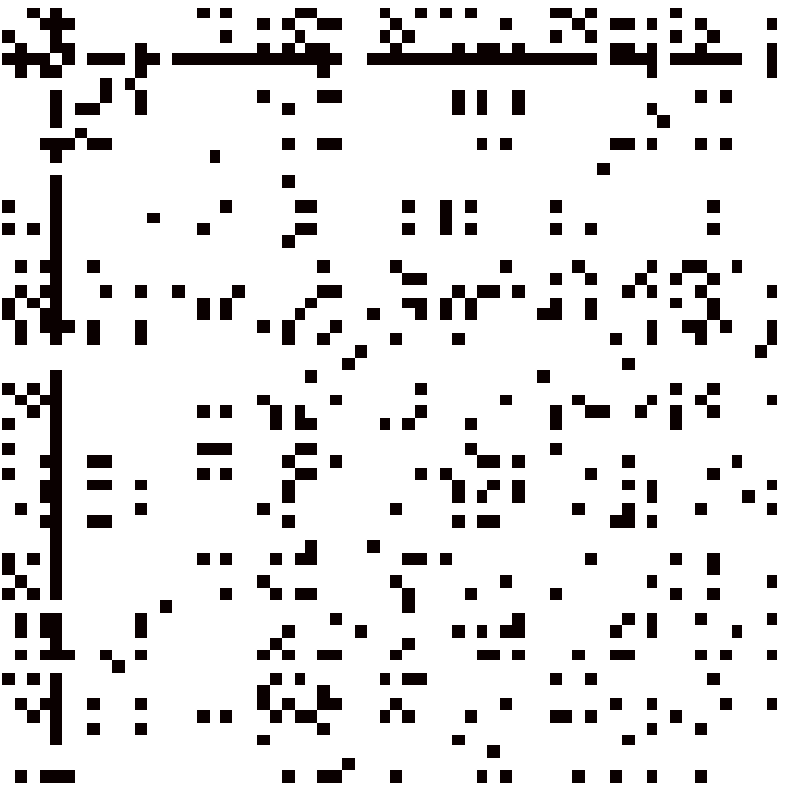}}}
\psline[linecolor=black, linewidth=0.04, arrowsize=0.05291667cm 2.0,arrowlength=1.4,arrowinset=0.0]{->}(5.22,-0.406521)(5.22,-2.006521)
\rput(0.02, -3){\fboxsep0pt\fbox{\includegraphics[scale=0.25]{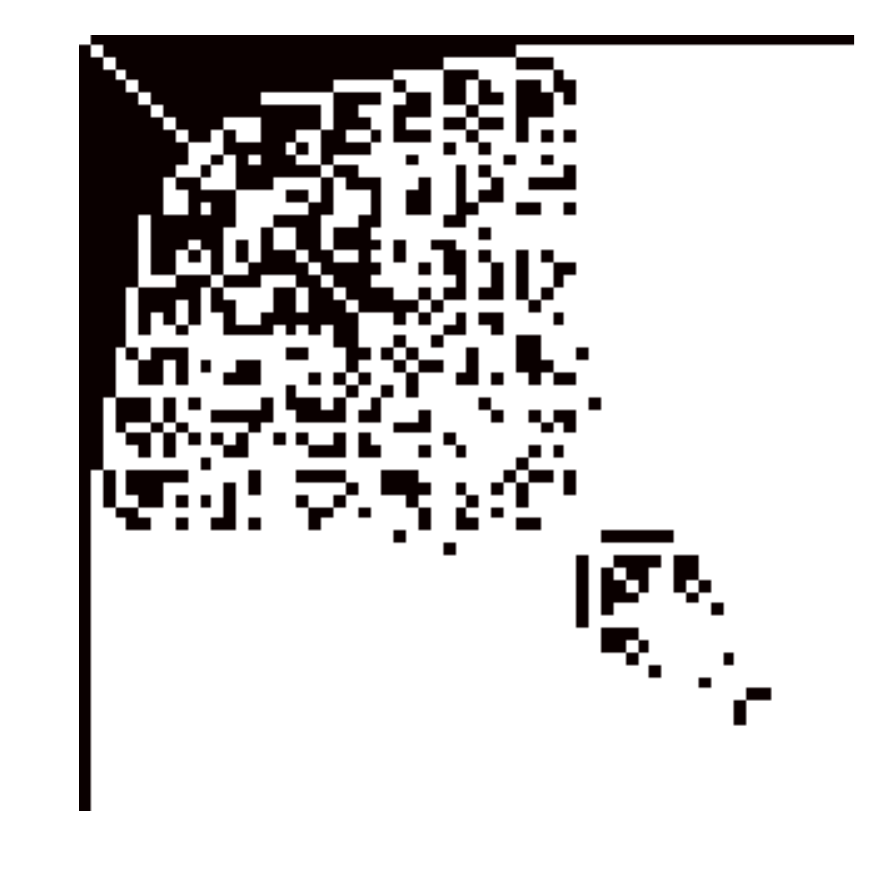}}}
\rput(5.22, -3){\fboxsep0pt\fbox{\includegraphics[scale=0.25]{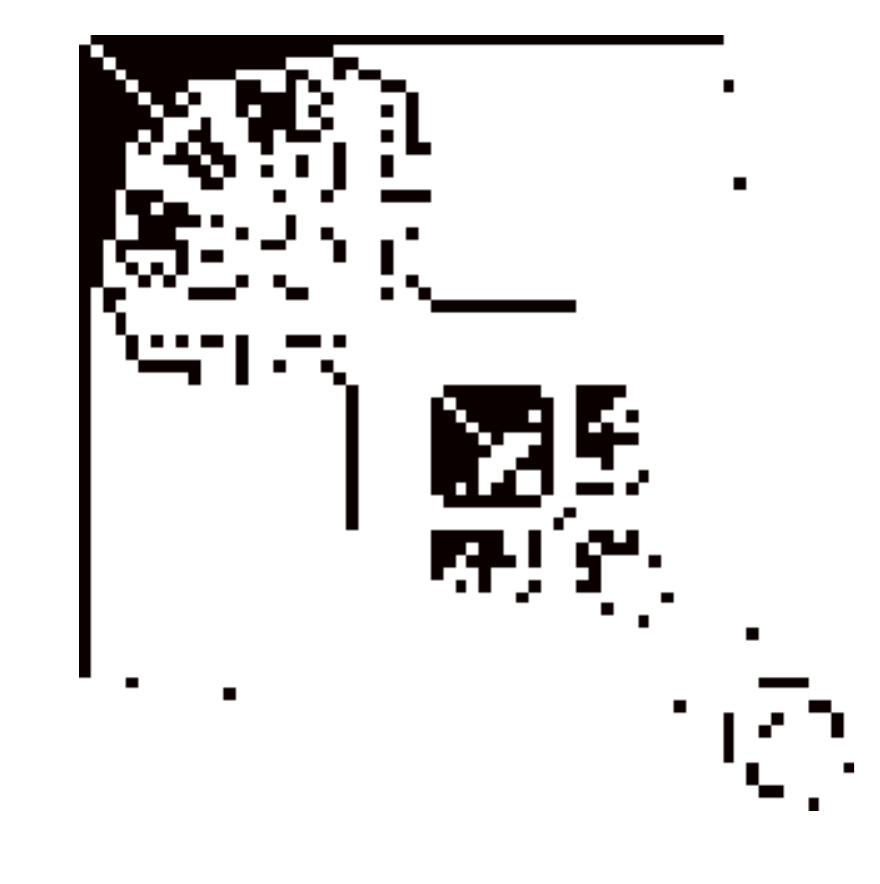}}}
\rput(2.50, 4.35){Subgraphs}
\rput(2.50, 0.75){\shortstack{Adjacency\\Matrices}}
\rput(2.50, -3){\shortstack{Structured Image\\Embeddings}}
\end{pspicture}
}

\caption{Obtaining structured image embeddings of graphs}
\label{fig:imem}
\end{figure}

We briefly describe the process of converting adjacency matrices to lossless image features \cite{wu2016asonam} here. The adjacency matrices can be visualized as images by simply treating 1s as black pixels and 0s as white pixels. However, the same adjacency matrix can be mapped to different images by permuting the rows. Using the image from a random permutation of the rows as input directly to a classifier such as a Convolutional Neural Network (CNN) results in very poor results. It is necessary to first re-order the nodes in a canonized form. We use the ordering scheme shown in \cite{wu2016asonam}, to make sure that all permutations of a given adjacency matrix map to the same structured image making it permutation invariant. When these structured images are fed to a CNN, classification performance is significantly improved. Neural networks show tremendous accuracy when it comes to recognizing real world images \cite{jia2014caffe}. As shown in \cite{hegdesig2018}, they do very well with homogenous networks as well. Different subgraphs from the same network are different at the microscopic level but are similar on a macroscopic level. We use the image embeddings of local subgraphs to identify the different parts of heterogeneous graphs. Figure \ref{fig:imem} is a visualization of the process.

\subsection{Pre-trained Models}
\label{sec:pretrain}
In \cite{hegdesig2018}, authors test several classifiers with the task of discriminating between the real world networks mentioned in Table \ref{tab:data}. CNN performs best with about 86\% accuracy. The model is trained on the subgraphs extracted from the homogenous networks separately and it learns each of the individual network's signature. We use the CNN model trained in this setting here. However, in our current setting the test data consists of snapshots taken from the heterogeneous network constructed as described in Section \ref{sec:consthn}. We use the model trained in the homogenous setting so that it has learned the \emph{pure} local structures. If trained in the heterogenous setting, the model will fail to effectively learn network signatures because of lack of consistent local structure.

\section{Experiments and Results}
\label{sec:exptres}

\subsection{Node Classification in Heterogenous Networks}
\label{sec:nodeclass}
We perform random walks starting from every node in the network. Then, we obtain the structured image embedding of each subgraph. This is fed as a test sample to the already trained CNN model (on the real world networks) to get a label. We assign this label to the starting node of the random walk and all the other nodes in the test subgraph. However, we maintain these two sets of labels for each node separately. One set has the label a node receives when it is the starting node in the random walk and the second set contains all the labels it receives when it is not. Finally, we repeat the process with random walk lengths (lenses) of 8, 16, 32 and 64. Thus, each node gets 8 sets of labels. We show the individual accuracies of each of the lenses in Table \ref{tab:accs}. One can see that lens sizes 16 and 32 perform better than the smallest lens (8) and the largest lens (64). This is because the smallest lens \emph{zooms} in too much into the network and the local signature is not captured optimally. Similarly, with the biggest lens, it looks at more than one local network signature in one snapshot which causes error.

\begin{table}[h]
\caption{Performances (percent correct) of different lenses. First column: predicted label is assigned to all nodes in the test subgraph. Second column: predicted label is assigned only to the starting node of the test subgraph.}
\label{tab:accs}
\centering
\resizebox{0.4\textwidth}{!}{%
\begin{tabular}{c|c|c}
\shortstack{Lens\\Sizes} & \shortstack{Label assigned\\to all nodes} & \shortstack{Label assigned to\\starting node only}\tabularnewline
\hline 
8 & 27.55 & 29.08\tabularnewline
16 & 32.94 & 33.96\tabularnewline
32 & 33.42 & 34.53\tabularnewline
64 & 30.64 & 30.65\tabularnewline
\end{tabular}}
\end{table}

Each node gets a set of labels from different lenses. Rather than just using one of the lenses as the final classifier, we can aggregate all the labels to get a more accurate classification that incorporates the information from all the lenses. To this end, we split the nodes into training and test sets and use the training set to \emph{learn} the optimal weights to weigh the label sets from each of the lenses. Consider the matrix $X_m \in \mathbb{R}^{8 \times 9}$ that is maintained for each of the $M$ nodes where the $ij^{th}$ entry denotes the number of times lens $i$ gave the label $j$ to node $m$.

\[ X_m = 
\begin{bmatrix}\ddots & \vdots & \iddots\\
\ldots & n_{ij} & \ldots\\
\iddots & \vdots & \ddots
\end{bmatrix}
\]

Now, we assign weights $p_i$ to the lenses such that the weighted sum of the column corresponding to the correct label is maximum. Let $y_m$ denote this column for node $x_m$. This condition can be written as:

\[
\mathbf{1} \cdot y^T_m \cdot p \geq X^T_m \cdot p
\]

To allow for error, we introduce slack variables $\xi_m$ and require $\mathbf{1} \cdot y^T_m \cdot p \geq X^T_m \cdot p - \xi_m$. The objective is to minimize the sum of errors which gives a linear program:

\begin{alignat*}{2}
  & \text{minimize: } & & \sum_{m} \xi_{m} \\
   & \text{subject to: }& \quad & \begin{aligned}[t]
                    (\mathbf{1} \cdot y^T_m - X^T_m) \cdot p & \geq -\xi_m,& \\
                  \sum_ip_i & = 1, & \quad \\
                  0 \leq p_i & \leq 1, & \quad \\
                  \xi_m & \geq 0
                \end{aligned}
\end{alignat*}

Alternatively, one can na\"ively assign the accuracy score of individual lenses as shown in Table \ref{tab:accs} as the weights to the corresponding lenses. By applying the weights, we obtain a probability distribution over the 9 possible labels for each node. We classify the node as the top-$k$ labels, where $k$ is chosen so that the sum of the probabilities exceeds a threshold $\tau$.

For example, let node $m$ have the following set of labels and associated probabilities: $\{\textrm{amazon}: 0.5, \textrm{ facebook}: 0.3, \textrm{ road}: 0.2\}$ and the rest 0s. If threshold $\tau$ is set to 0.8, the set would be trimmed down to just $\{\textrm{amazon}: 0.5, \textrm{ facebook}: 0.3\}$. For a given threshold, for each node we calculate the classification accuracy as:

\[
\textrm{accuracy} = 
\begin{cases}
1/k & \textrm{if true label} \in \textrm{top-}k(\tau)\\
0 & \textrm{otherwise}
\end{cases}
\]

\begin{figure}
\hspace{-15pt}
\includegraphics[scale=0.75]{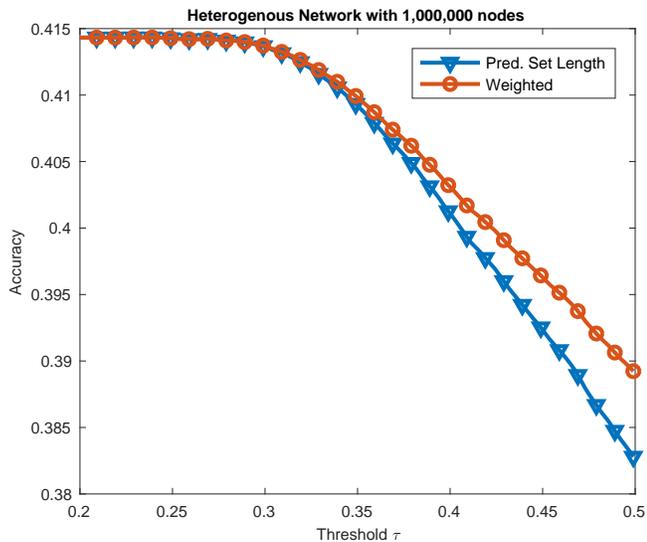}
\caption{Performance of the node classification model with lenses weighted by optimal weights obtained by the linear program on the larger million node heterogenous network}
\label{fig:linprogscore1000}
\end{figure}

The final reported accuracy score is the average accuracy over all test nodes. We show accuracies as a function of $\tau$ for the weighs obtained from linear programming in Figure \ref{fig:linprogscore1000}. In general, small values $\tau$ which result in a small final label set size is best. The accuracy is about 42\%.

We present the confusion matrix at peak $\tau$ of this analysis in Table \ref{tab:rawconfhetero}. Each entry $(i,j)$ in Table \ref{tab:rawconfhetero} represents the total reward received for all nodes of type $i$ in the heterogeneous network for being type $j$. When $i=j$, the reward represents correct classification. The sum of the diagonals of the confusion matrix divided by the sum of all the entries yields the accuracy of the model. 



\subsection{Diversity of Nodes' Network Connections}
\label{sec:nodeloc}
\begin{figure*}
\begin{center}
\begin{subfigure}{.5\textwidth}
\begin{center}
\includegraphics[scale=0.6]{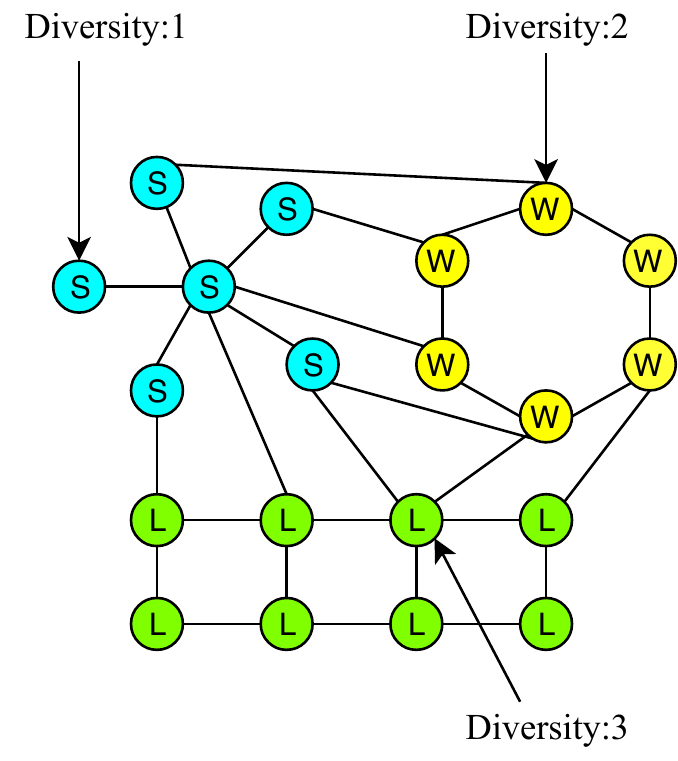}
\end{center} 
\caption{\shortstack{The number of different types of networks a node is\\connected to is referred to as its node diversity.}}
\label{fig:diversitypic}
\end{subfigure}%
\begin{subfigure}{.5\textwidth}
\vspace{20pt}
\hspace*{-17pt}
\begin{tabular}{c|c|c|c}
\shortstack{Node\\Diversity} & \shortstack{Top Label\\Correct (\%)} & \shortstack{Average Top\\Weight} & \shortstack{Average Entropy of\\Weights (normalized)}\tabularnewline
\hline 
1 & 50.39 & 0.53 & 0.57\tabularnewline
2 & 32.69 & 0.47 & 0.63\tabularnewline
3 & 19.85 & 0.44 & 0.67\tabularnewline
4 & 13.00 & 0.42 & 0.69\tabularnewline
5 & 12.30 & 0.41 & 0.70\tabularnewline
\hline 
\multicolumn{2}{c|}{\shortstack{Correlation with\\Node Diversity {[}-1, 1{]}}} & -0.1867 & +0.2466\tabularnewline
\end{tabular}
\vspace{15pt}
\caption{\shortstack{Top weight decreases and entropy of weights increases\\as node diversity increases.}}
\label{tab:loc}
\end{subfigure}
\caption{Node diversity - as captured by our Network Lens.}
\label{fig:diversity}
\end{center}
\end{figure*}

The higher the number of different networks a node is connected to, lesser the chance that the label with the maximum weight is correct. We can see in Figure \ref{fig:diversity}.

Note that even though a node could be connected to just one type of network, its neighbors could be from a different sized subgraph. For example, consider a `Facebook' node in a 32-node subgraph connected to another `Facebook' node in a 16-node subgraph. This node is still connected to only 1 type of network. Since the lenses are trained in a setting where all the subgraphs are of the same size, the previously mentioned scenario can cause error.

The reason why diverse nodes fare worse is because the random walk starting from a diverse node can meander into different network types. Thus, the subgraph collected from a diverse node will produce a structured image embedding which is corrupted by the different networks, increasing the chances of a wrong prediction.

If $w_i$ are the weights in the predicted label set for node $i$, then entropy is given by $-\sum_i w_i log(w_i)$. We calculate the entropy for each node and report the normalized average along with the average top label weight in Table \ref{tab:loc}. We see that there is high correlation between the diversity of the nodes and their top label weight as well as with the entropy of their weights. As node diversity increases, top label weight decreases (the less has less confidence in the top label) and the weight entropy increases (the confidence of the lens is more spread out among the labels).

\subsection{Homogeneity in Networks}
\label{sec:homonw}
We repeat the experiments in Section \ref{sec:nodeclass} but with each of the individual networks that were spliced together to form the big heterogenous network. The purpose of these experiments is to study how homogenous real networks are, for example, how much of the `Facebook' network, actually behaves like a `Facebook' network.

\begin{figure}
\centering
\includegraphics[scale=0.7]{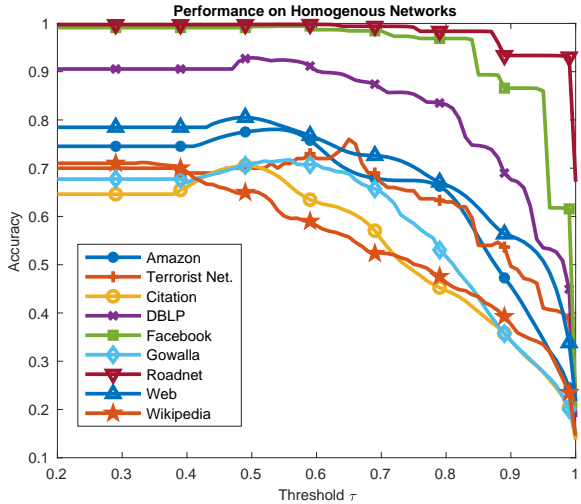}
\caption{Performance on homogenous networks}
\label{fig:homo}
\end{figure}

First, we look at Table \ref{tab:homo} which shows test performance on each of the networks. The networks with high accuracy like Road Network, Facebook and DBLP have high homogeneity. We also show the mode of the incorrect labels for each network. This is the label that was assigned to a network the most times other than the correct label. It is interesting that 5 out of the 9 networks got labeled as Citation most frequently. Since Citation has the poorest performance this shows that it is the least homogeneous of all the networks.

\begin{table}
\centering
\resizebox{0.4\textwidth}{!}{%
\begin{tabular}{c|c|c}
\shortstack{Network\\ \hspace{1pt}} & \shortstack{Peak Acc.\\(\%)} & \shortstack{Mode of\\Incorrect Labels}\tabularnewline
\hline 
Road Net. & 99.80 & Citation\tabularnewline
Facebook & 99.44 & Web\tabularnewline
DBLP & 92.88 & Citation\tabularnewline
Web & 80.51 & Amazon\tabularnewline
Amazon & 78.05 & Citation\tabularnewline
Terrorist Net. & 76.00 & Gowalla\tabularnewline
Gowalla & 71.71 & Citation\tabularnewline
Wikipedia & 71.19 & Citation\tabularnewline
Citation & 70.57 & Amazon\tabularnewline
\end{tabular}}
\caption{Homogeneity in Networks}
\label{tab:homo}
\end{table}

In Figure \ref{fig:homo}, we observe that networks like Amazon and Wikipedia do have some heterogeneity compared to, say, Facebook and the road network. This is because, the label with the top weight for a node in the Amazon network only has about 78.05\% accuracy compared to near 100\% accuracy with Facebook. This shows that the Amazon network has less inherent structure where as Facebook has high inherent structure and hence is more homogenous.

In Table \ref{tab:rawconf} we present the raw numbers behind the homogeneity analysis. Table \ref{tab:rawconf} must be read row-wise. Entry $j$ in row $i$ represents the total reward the nodes of network $i$ received for being type $j$. When $i=j$, the reward represents correct classification. Note that this is not a confusion matrix, but can be thought of as a confusion ``row". This is because there is only one class in each of the test sets since networks are tested one at a time. Also, every row sums to a different number since each network is of a different size resulting in different sized test sets (20\%). The optimal weights are learned from the training set and their performance on the test set is shown in Figure \ref{fig:homo}.

\begin{table*}
\centering
\caption{Number of different labels assigned to each type of node in the million node heterogenous network by all the lenses}
\label{tab:rawconfhetero}
\begin{tabular}{c|ccccccccc}
 & Amazon & Terrorist Net. & Citation & DBLP & Facebook & Gowalla & Road Net. & Web & Wikipedia\tabularnewline
\hline 
Amazon & \textbf{17377.5} & \textcolor{gray}{20.5} & \textcolor{gray}{1885.5} & \textcolor{gray}{2026.5} & \textcolor{gray}{1034.3} & \textcolor{gray}{196.5} & \textcolor{gray}{444.5} & \textcolor{gray}{816} & \textcolor{gray}{98.5}\tabularnewline
Terrorist Net. & \textcolor{gray}{8423} & \textbf{7107.5} & \textcolor{gray}{935} & \textcolor{gray}{4752} & \textcolor{gray}{1151.5} & \textcolor{gray}{1080} & \textcolor{gray}{115.5} & \textcolor{gray}{366.5} & \textcolor{gray}{10}\tabularnewline
Citation & \textcolor{gray}{9424.5} & \textcolor{gray}{15.5} & \textbf{9283.5} & \textcolor{gray}{1737} & \textcolor{gray}{1892.5} & \textcolor{gray}{378.5} & \textcolor{gray}{469.5} & \textcolor{gray}{433} & \textcolor{gray}{265}\tabularnewline
DBLP & \textcolor{gray}{9099} & \textcolor{gray}{22} & \textcolor{gray}{1214} & \textbf{11155} & \textcolor{gray}{1220} & \textcolor{gray}{143} & \textcolor{gray}{360} & \textcolor{gray}{659.3} & \textcolor{gray}{46}\tabularnewline
Facebook & \textcolor{gray}{4398} & \textcolor{gray}{6.5} & \textcolor{gray}{710.5} & \textcolor{gray}{1530} & \textbf{16451} & \textcolor{gray}{130} & \textcolor{gray}{74.50} & \textcolor{gray}{545} & \textcolor{gray}{10.5}\tabularnewline
Gowalla & \textcolor{gray}{9622.3} & \textcolor{gray}{83} & \textcolor{gray}{2163.5} & \textcolor{gray}{3394} & \textcolor{gray}{1456} & \textbf{4201} & \textcolor{gray}{128} & \textcolor{gray}{2497.5} & \textcolor{gray}{338.5}\tabularnewline
Road Net. & \textcolor{gray}{9517} & \textcolor{gray}{16.5} & \textcolor{gray}{1776.7} & \textcolor{gray}{1843} & \textcolor{gray}{1707} & \textcolor{gray}{280} & \textbf{7854} & \textcolor{gray}{777} & \textcolor{gray}{43}\tabularnewline
Web & \textcolor{gray}{7517.8} & \textcolor{gray}{7} & \textcolor{gray}{809} & \textcolor{gray}{2208.5} & \textcolor{gray}{779.5} & \textcolor{gray}{292.5} & \textcolor{gray}{706.5} & \textbf{11505.3} & \textcolor{gray}{72}\tabularnewline
Wikipedia & \textcolor{gray}{9615.3} & \textcolor{gray}{15} & \textcolor{gray}{5342.5} & \textcolor{gray}{1105} & \textcolor{gray}{987} & \textcolor{gray}{1588.5} & \textcolor{gray}{495.5} & \textcolor{gray}{530.5} & \textbf{4188.5}\tabularnewline
\end{tabular}
\end{table*}

\begin{table*}
\centering
\caption{Number of different labels assigned to the nodes of each of the homogeneous networks by all the lenses}
\label{tab:rawconf}
 \begin{tabular}{c|ccccccccc}
 & Amazon & Terrorist Net. & Citation & DBLP & Facebook & Gowalla & Road Net. & Web & Wikipedia\tabularnewline
\hline 
Amazon & \textbf{51262} & \textcolor{gray}{199} & \textcolor{gray}{4349.5} & \textcolor{gray}{1400} & \textcolor{gray}{112.5} & \textcolor{gray}{2291} & \textcolor{gray}{2121} & \textcolor{gray}{2250} & \textcolor{gray}{2987}\tabularnewline
Terrorist Net. & \textcolor{gray}{0} & \textbf{41} & \textcolor{gray}{0} & \textcolor{gray}{0} & \textcolor{gray}{2} & \textcolor{gray}{11} & \textcolor{gray}{0} & \textcolor{gray}{0} & \textcolor{gray}{0}\tabularnewline
Citation & \textcolor{gray}{1162} & \textcolor{gray}{1.5} & \textbf{4871} & \textcolor{gray}{96.5} & \textcolor{gray}{82.5} & \textcolor{gray}{390} & \textcolor{gray}{12} & \textcolor{gray}{47.5} & \textcolor{gray}{246}\tabularnewline
DBLP & \textcolor{gray}{2368.5} & \textcolor{gray}{59} & \textcolor{gray}{966} & \textbf{58900.5} & \textcolor{gray}{45.5} & \textcolor{gray}{430} & \textcolor{gray}{4} & \textcolor{gray}{581.5} & \textcolor{gray}{61}\tabularnewline
Facebook & \textcolor{gray}{2} & \textcolor{gray}{1} & \textcolor{gray}{1} & \textcolor{gray}{1} & \textbf{801} & \textcolor{gray}{1} & \textcolor{gray}{0} & \textcolor{gray}{1} & \textcolor{gray}{0}\tabularnewline
Gowalla & \textcolor{gray}{3392} & \textcolor{gray}{4.5} & \textcolor{gray}{4697} & \textcolor{gray}{629} & \textcolor{gray}{142} & \textbf{27835.3} & \textcolor{gray}{256.2} & \textcolor{gray}{1728.2} & \textcolor{gray}{633.8}\tabularnewline
Road Net. & \textcolor{gray}{173} & \textcolor{gray}{3} & \textcolor{gray}{320} & \textcolor{gray}{73} & \textcolor{gray}{0} & \textcolor{gray}{0} & \textbf{216996} & \textcolor{gray}{3} & \textcolor{gray}{50}\tabularnewline
Web & \textcolor{gray}{12109} & \textcolor{gray}{453} & \textcolor{gray}{4774.5} & \textcolor{gray}{7775} & \textcolor{gray}{3179.5} & \textcolor{gray}{443} & \textcolor{gray}{646} & \textbf{140672.5} & \textcolor{gray}{5090.5}\tabularnewline
Wikipedia & \textcolor{gray}{24} & \textcolor{gray}{1} & \textcolor{gray}{226} & \textcolor{gray}{0.50} & \textcolor{gray}{0.5} & \textcolor{gray}{4.5} & \textcolor{gray}{0} & \textcolor{gray}{8} & \textbf{656.5}\tabularnewline
\end{tabular}
\end{table*}


%

\section{Conclusion and Future Work}
\label{sec:concfw}
In summary, we successfully used a new way to classify a small subnetwork of a topologically heterogeneous network using structured image embeddings. We showed that this technique is highly scalable since we achieved high accuracies on a million node network. We believe our simple and easy to understand model coupled with its strong performance will pave the way for new research and applications. A future direction is to increase the database of pre-trained classifiers to improve the diversity.

\section*{Acknowledgment}
\label{sec:ack}
This research was supported by the Army Research Laboratory under Cooperative Agreement W911NF-09-2-0053 (the ARL-NSCTA). The views and conclusions contained in this document are those of the authors and should not be interpreted as representing the official policies, either expressed or implied, of the Army Research Laboratory or the U.S. Government. The U.S. Government is authorized to reproduce and distribute reprints for government purposes notwithstanding any copyright notation here on.

\bibliography{example_paper}
\bibliographystyle{icml2018}
\end{document}